\documentclass[aps,prl,nofootinbib,twocolumn,superscriptaddress,preprintnumbers]{revtex4-1}
\usepackage[nodisplayskipstretch]{setspace}
\usepackage{cancel}

\usepackage{amssymb}
\usepackage{amsmath}
\usepackage{epsfig}
\usepackage{hyperref}
\usepackage{breakurl}
\usepackage{bm}
\usepackage{color}
\usepackage{tikz}
\usetikzlibrary{decorations.markings}
\usepackage[english]{babel}
\newcommand\beq{\begin{equation}}
\newcommand\eeq{\end{equation}}
\def\bea{\begin{eqnarray}}
\def\eea{\end{eqnarray}}

\DeclareRobustCommand{\SkipTocEntry}[4]{}

\DeclareMathOperator{\Arcsinh}{sinh^{-1}}

\newcommand{\nn}{\nonumber}

\newcommand\beal{\begin{aligned}}
\newcommand\eeal{\end{aligned}}

\newcommand\dd{{\mathrm d}}
\usepackage{xcolor}

\newcommand{\bk}{{\boldsymbol k}}

\newcommand{\bp}{{\boldsymbol p}}

\newcommand{\bq}{{\boldsymbol q}}
\newcommand{\br}{{\boldsymbol r}}

\newcommand\Mp{M_{\rm Pl}}


\begin{document}

\preprint{\tt DESY 20-114} 
\preprint{\tt SLAC-PUB-17545}

\title{Conservative Dynamics of Binary Systems to Third Post-Minkowskian Order\\ [0.1cm] from the Effective Field Theory Approach}
\author{Gregor K\"alin}
\affiliation{SLAC National Accelerator Laboratory, Stanford University, Stanford, CA 94309, USA}
\author{Zhengwen Liu}
\author{Rafael A. Porto}
\affiliation{ Deutsches Elektronen-Synchrotron DESY,
Notkestrasse 85, 22607 Hamburg, Germany}

\begin{abstract}
We derive the conservative dynamics of non-spinning binaries to third~Post-Minkowskian order, using the Effective Field Theory (EFT) approach introduced in \cite{pmeft} together with the Boundary-to-Bound dictionary developed in \cite{paper1,paper2}. The main ingredient is the scattering angle, which we compute to ${\cal O}(G^3)$ via Feynman diagrams. Adapting to the EFT framework powerful tools from the amplitudes program, we show how the associated (master) integrals are {\it bootstrapped} to all orders in velocities via differential equations. Remarkably, the boundary conditions can be reduced to the same integrals that appear in the EFT with Post-Newtonian sources. For the sake of comparison, we reconstruct the Hamiltonian and the classical limit of the scattering amplitude. Our~results are in perfect agreement with those~in Bern~et~al.~\cite{zvi1,zvi2}. 
\end{abstract}


\maketitle

\emph{Introduction.} The discovery potential heralded by the new era of gravitational wave (GW) science \cite{LIGOScientific:2018mvr,LIGO}  has motivated high-accuracy theoretical predictions for the dynamics of binary systems~\cite{buosathya,tune,music}. This is particularly important for the inspiral phase of small relative velocities ($v/c \ll 1$), covering a large portion of the cycles in the detectors' band for many events of interest, which is amenable to perturbative treatments like the celebrated Post-Newtonian (PN) expansion \cite{blanchet, Schafer:2018kuf}. Notably, in parallel with more `traditional' approaches in general relativity, e.g.~\cite{Damour:2014jta,Jaranowski:2015lha,Bernard:2015njp,Bernard:2017bvn,Marchand:2017pir}, in recent years ideas from particle physics, such as Effective Field Theories (EFTs) similar to those used to study bound states of strongly interacting particles \cite{nrgr,walterLH,foffa,iragrg,grg13,review}, and modern tools from scattering amplitudes connecting gravity to Yang-Mills theory and bypassing Feynman diagrams \cite{elvang,reviewdc}, have found their way into the classical two-body problem in gravity. Although more recent, these novel tools have made~key contribution to the knowledge of the conservative dynamics of binary systems, both in the PN regime as well as the Post-Minkowskian (PM) expansion in powers of $G$ (Newton's constant), with the present state-of-the-art reaching the fourth PN (4PN) \cite{nrgr2pn,nrgr3pn,Foffa:2012rn,tail,nrgrG5,apparent,nrgr4pn1,nrgr4pn2} and third PM (3PM) \cite{cheung,zvi1,zvi2} orders for non-spinning bodies, respectively.\footnote{ Partial results are also known to 5PN (static)~\cite{5pn1,5pn2} and 6PN \cite{blum,bini2}; radiation and spin are incorporated in e.g \cite{dis1,andirad,andirad2,adamchad1,radnrgr,nrgrs,prl,Porto:2007tt,dis2,nrgrss,nrgrs2,nrgrso,rads1,amps,maiaso,maiass,levi,Levi:2020kvb,Levi:2020uwu,Vaidya:2014kza,Guevara:2019fsj,Arkani-Hamed:2019ymq,Chung:2020rrz,Bern:2020buy}.} \vskip 4pt

Gravitational scattering amplitudes \cite{cheung,zvi1,zvi2} find a natural~habitat in the PM regime of a quantum world, which, at first, appears to bear little connection to the classical bound states where traditional PN tools \cite{blanchet} and~EFT approach \cite{review}  have been applied so far. While this can be circumvented by the universal character of the interaction, which is independent of the state, one still~has to extract the classical part of the amplitude. In the framework of \cite{zvi1,zvi2,cheung}, this relies on the large angular momentum limit $\tfrac{\hbar}{J} \to 0$ (resulting also in a series of spurious infrared divergences removed by a matching computation). The procedure, however, was challenged in~\cite{damour3}, with doubts (some addressed~in~\cite{blum,bini2}) on the validity of~the~3PM Hamiltonian in \cite{zvi1,zvi2}. In~light of its relevance, and demand for even higher accuracy~\cite{Antonelli:2019ytb}, a systematic, scaleable, and purely classical approach to observables in the PM regime was thus imperative. \vskip 4pt

Building~upon the universal {\em boundary-to-bound} (B2B) dictionary, relating scattering data directly to gauge-invariant observables for generic orbits through analytic continuation \cite{paper1,paper2}, a novel PM framework was developed in \cite{pmeft} using the EFT machinery, and readily implemented for bound states to ${\cal O}(G^2)$. (See e.g.~\cite{damour1,damour2,Antonelli:2019ytb,damour3n,damour3,eob} for alternative routes.) In this letter we report the next step in the EFT approach, namely the computation of the conservative binary dynamics to 3PM order. This entails the calculation of the scattering~angle to next-to-next-to-leading order (NNLO) in~$G$ via Feynman diagrams. Remarkably, we find that the associated (master) integrals can be {\it bootstrapped} from their PN counterparts through differential equations in the velocity \cite{Henn:2014qga}, as advocated in~\cite{Parra}, paving the way forward to higher order computations. For the sake of comparison, we reconstruct the Hamiltonian as well as the (infrared-finite) amplitude in the classical limit, and find complete agreement with the results in~\cite{zvi1,zvi2}. Our derivation thus independently confirms the connection between the amplitude and the center-of-mass (CoM) momentum ({\em impetus formula}) \cite{paper1}, and the legitimacy of the program to extract classical physics from scattering amplitudes \cite{paper1,paper2, ira1,cheung,zvi1,zvi2,donal,donalvines, withchad,Holstein:2008sx,Bjerrum-Bohr:2013bxa,Vaidya:2014kza,Guevara:2017csg,Chung:2018kqs,Guevara:2018wpp,Walter,simon,Guevara:2019fsj,bohr,cristof1,Arkani-Hamed:2019ymq,Bjerrum-Bohr:2019kec,Chung:2019duq,Bautista:2019tdr,Bautista:2019evw,KoemansCollado:2019ggb,Brandhuber:2019qpg,Johansson:2019dnu,Aoude:2020onz,Cristofoli:2020uzm,Chung:2020rrz,Bern:2020gjj,Bern:2020buy,Cheung:2020gyp,Parra,soloncheung,AccettulliHuber:2020oou}. At the same time, we explicitly demonstrate the power of the EFT and B2B framework \cite{pmeft,paper1,paper2}, which by design can be systematized to all orders. 
\vskip 2pt
\begin{figure}[t!] 
\includegraphics[width=0.25\textwidth]{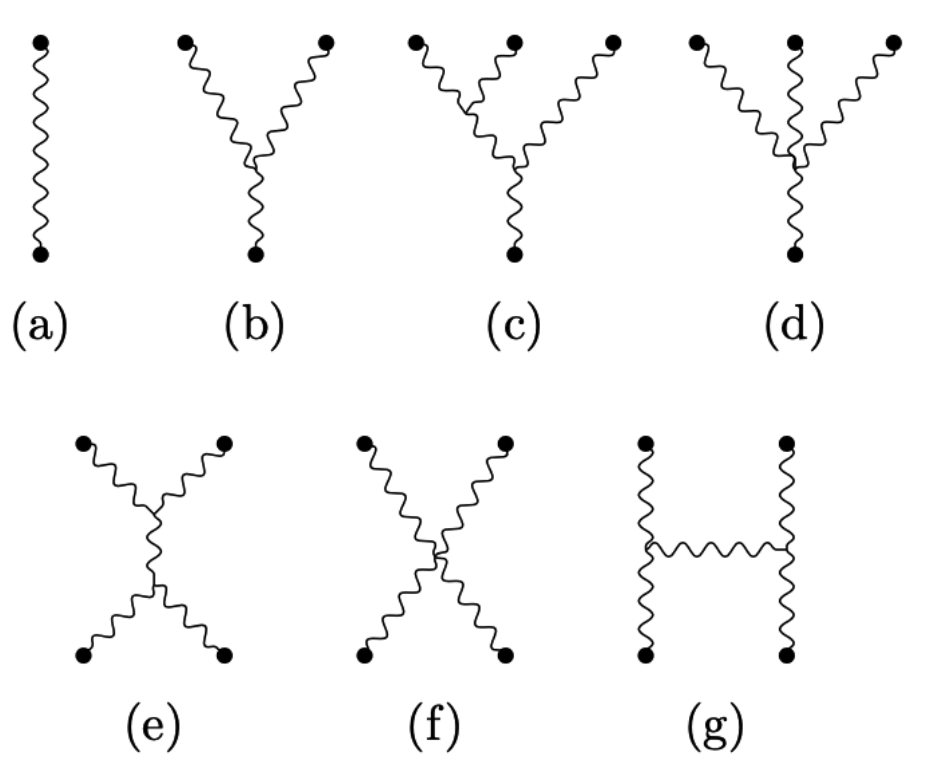} 
      \caption{Feynman topologies to 3PM~\cite{pmeft}.} 
      \label{fig1}
            \vspace{-0.4cm}
\end{figure}

\emph{The EFT framework.} The starting point is the effective action from which we derive the scattering trajectories. We proceed by \emph{integrating out} the metric field $g_{\mu\nu} = \eta_{\mu\nu}+ h_{\mu\nu}/\Mp$ (with $\Mp^{-1} \equiv \sqrt{32\pi G}$)
\beq
e^{i S_{\rm eff} } = \int {\cal D} h_{\mu\nu} \, e^{i S_{\rm EH}[h] + i S_{\rm GF}[h] + i S_{\rm pp}[x_a,h]}\,,\label{eff1}
\eeq
in the (classical) saddle-point and weak-field approximations. We~work with the Einstein-Hilbert~action,~$S_{\rm EH}$, and the convention $\eta_{\mu\nu}= {\rm diag}(+,-,-,-)$. The~gauge-fixing, $S_{\rm GF}$, is adjusted to simplify the Feynman rules~\cite{pmeft}. We~use the (Polyakov) point-particle effective~action, \beq
S_{\rm pp} = -\sum_{a = 1,2} \frac{m_a}{2} \int \dd\tau_a\,g_{\mu\nu}(x^\alpha_a) v_a^\mu v_a^\nu + \cdots\,,\label{act}
\eeq
with $\tau_a$ the proper time. The ellipses include higher-derivative terms accounting for finite-size effects~and counterterms to remove (classical) ultraviolet divergences \cite{nrgr,pmeft}. As~usual, we use dimensional~regularization.\vskip 2pt

\emph{Impulse from Action.} From the action we read~off~the effective Lagrangian at each order in $G$: ${\cal L}_{\rm eff} = {\cal L}_{0} + {\cal L}_1 +{\cal L}_2 + {\cal L}_3+ \cdots $. Although it may be non-local in time when radiation-reaction effects are included \cite{Damour:2014jta,tail}, it~is manifestly local with only potential modes \cite{pmeft}. Using~the effective Lagrangian we obtain the trajectories,
\beq
\label{pmexp}
\begin{aligned}
x^\mu_a(\tau_a) &= b^\mu_a + u^\mu_a \tau_a + \sum_n \delta^{(n)} x^\mu_a (\tau_a)\,,
\end{aligned}
\eeq
with $u^\mu_a$ the velocity at infinity, obeying $u_a^2=1$, and $b^\mu \equiv b^\mu_1 - b_2^\mu$ the impact parameter. For instance, at LO, 
\begin{align}\label{d1x}
\delta^{(1)} x^\mu_1 (\tau_1) =  - &\frac{m_2}{8\Mp^2}  \big( (2\gamma^2 - 1) \eta^{\mu\nu} - 2(2\gamma u_2^\mu - u_1^\mu) u_1^\nu \big) \nonumber\\
& \times \int_k\frac{ik_\nu\,\hat\delta(k\cdot u_2)\, e^{i k\cdot b}}{k^2(k\cdot u_1-i0^+)^2}e^{i (k\cdot u_1-i0^+) \tau_1}.
\end{align}
We use the notation $\int_k \equiv \int \frac{\dd^4k}{(2\pi)^4}$, $\hat\delta(x) \equiv 2\pi\delta(x)$~and 
\beq \gamma \equiv u_1\cdot u_2 = \frac{p_1\cdot p_2}{m_1m_2} = \frac{E_1E_2+\bp^2}{m_1m_2}\,,\eeq where $E_a = \sqrt{\bp^2+m_a^2}$ and $\pm\bp$ is the incoming CoM momentum. Notice the factor of $(k\cdot u_1-i0^+)^{-1}$, with the $i0^+$ to ensure convergence of the time integrals, which resembles the linear propagators appearing in heavy-quark effective theory \cite{ben}. The pole shifts to $(k\cdot u_2+i0^+)^{-1}$ for particle~2. The impulse follows from the effective action,
\beq
\Delta p_a^\mu = -\eta^{\mu\nu} \int_{-\infty}^{+\infty} \dd\tau_a \frac{\partial {\cal L_{\rm eff}}}{\partial x^\nu_a}(x_a(\tau_a))\,,\label{dp}
\eeq
where the overall sign is due to our conventions. The impulse can then be solved iteratively, starting with the undeflected trajectory in \eqref{pmexp}. Notice that all of the ${\cal L}_{k< n}$'s contribute to $n$PM order, and must be evaluated on the trajectories up to $(n{-}k)$-th order~in~$G$. We refer to this procedure as {\it iterations} \cite{pmeft}. The scattering angle, \beq
\frac{\chi}{2} = \sum_n \chi^{(n)}_b\left(\frac{GM}{b}\right)^n = \sum_n \frac{\chi^{(n)}_j}{j^n}\,,
\eeq 
with $1/j = GM\mu/(p_\infty b)$, is obtained from the relation
\beq
2\sin \frac{\chi}{2} = 2\left(\frac{\chi}{2} -\frac{1}{6} \left(\frac{\chi}{2}\right)^3 +\cdots\right) = \frac{\sqrt{-\Delta p_a^2}}{p_\infty}\,,\label{chi2}
\eeq
where \beq p_\infty = \mu \frac{\sqrt{\gamma^2-1}}{\Gamma}\,,\,\, \Gamma \equiv \frac{E}{M} = \sqrt{1+ 2\nu (\gamma-1)}\,,\eeq
with $E, M$ the total mass and energy, respectively. We use the notation $\mu = m_1m_2/M$ for the reduced mass, and $\nu = \mu/M$ for the symmetric mass ratio.\vskip 2pt The impulse may be further split into a contribution along the direction of the impact parameter as well as a term proportional to the velocities \cite{pmeft}. Due~to momentum conservation and the on-shell condition, we have \beq  (p_a+\Delta p_a)^2 = p_a^2 \,\, \Longrightarrow \, \, 2 p_a\cdot \Delta p_a = -\Delta p_a^2 \label{onshell}.\eeq  Moreover, since $\Delta^{(1)} p_1^\mu \propto b^\mu$ at leading PM order \cite{pmeft}, and $b\cdot u_a=0$, we can use \eqref{onshell} to solve iteratively for the component along the velocities. This allows us to restrict the derivation of the impulse to the perpendicular~plane~\cite{pmeft}. \vskip 2pt
\emph{Feynman Integrals.} To 3PM order the Feynman topologies are shown in~Fig.~\ref{fig1}. The computation yields four-dimensional relativistic integrals constrained by a series of $\delta$-functions, $\delta(k_i\cdot u_a)$, which arise due to the time integration in \eqref{dp} after inputting~\eqref{pmexp}. 
Moreover,  in addition to the standard factors of $1/k^2$ from the gravitational field, we have linear propagators, as in \eqref{d1x}, which are needed to compute the iterations. As we mentioned, we restrict ourselves to the computation of the impulse in the direction of the impact parameter. The derivation is then reduced to a series of terms proportional to the Fourier transform in the `transfer momentum',
\beq
 \int_q \hat\delta(q\cdot u_1) \hat\delta(q\cdot u_2)\, i q^\mu\, t^s\, M^{(a,\tilde a)}_{n_1n_2;i_1\cdots i_5} (q,\gamma)  e^{iq\cdot b}\,,\label{FT}
\eeq
where the factor of $t^s$, with $t\equiv -q^2$, depends on the tensor reduction of the given diagram. We find the following (cut) `two loop' integrals \cite{boot}
\beq
 M^{(a,\tilde a)}_{n_1n_2;i_1\cdots i_5}(q,\gamma) \equiv \int_{k_1,k_2} \frac{\hat\delta(k_1\cdot u_a)\hat\delta(k_2\cdot u_{\tilde a})}{A_{1,\not a}^{n_1} A_{2,\not\tilde a}^{n_2}\, D_1^{i_1} \cdots D_5^{i_5}}\,,\label{master}
\eeq
are sufficient to 3PM order,~where ($\not 1= 2$, $\not 2= 1$)
\begin{align}
\begin{aligned}
&A_{1,\not a} = k_{1}\cdot u_{\not a},\, A_{2,\not \tilde a}= k_2\cdot u_{\not{\tilde a}},\, D_1 = k_1^2,\, D_2=k_2^2\,, \\[0.35 em]
&D_3 = (k_1 {+} k_2 {-} q)^2,\, D_4 = (k_1 {-} q)^2,\, D_5=(k_2 {-} q)^2.
\end{aligned}
\end{align}
All the integrals we encounter in our computation, including the iterations, can be embedded into the family in \eqref{master} with different choices of $(a,\tilde a)$. The $i0$-prescription is such the $u_{1,2}$ are always accompanied by $\mp i0^+$, as~in~\eqref{d1x}. The other cases are obtained by different symmetrizations~\cite{boot}. We~keep only non-analytic terms in $t$ which yield long-range interactions~\cite{pmeft}. We outline the integration procedure momentarily. The outcome is the scaling  \beq t^s M^{(a,\tilde a)}_{n_1n_2; i_1 \cdots i_5}  \,\propto\,  {1 \over \epsilon}\,t^{-2\epsilon}\,, \eeq  with $\epsilon = (4-D)/2$, which gives for the impulse in \eqref{FT} the expected $b^\mu/b^4$ in $D=4$. The poles (and $\log\bar\mu$'s) in dimensional regularization accompanying the $\log t$'s produce contact terms that neatly drop out without referring to subtraction schemes~\cite{pmeft}.

 \vskip 4pt 
{\it Potential Modes.} In the framework of the PN expansion, the integrals would be performed using a mode factorization into {\it potential}  $(k_0 \ll |\bk|)$ and {\it radiation} $(k_0 \sim| \bk|)$ modes, while keeping manifest power counting in the velocity \cite{nrgr,Beneke:1997zp,review}. The computation with potential modes then reduces to a series of three-dimensional (massless) integrals. In contrast, in the PM scheme we ought to keep the propagators fully relativistic. The associated Feynman integral still receive contributions from both potential and radiation modes (yielding real and imaginary parts). We are interested here in the conservative sector, and we ignore for now radiation-reaction effects.\footnote{Hereditary tail effects, which enter in the conservative dynamics through a non-local contributions to the effective action e.g. \cite{Damour:2014jta,tail}, first appear at ${\cal O}(G^2a^2v^2) \sim {\cal O}(G^4 v^2)$ \cite{nrgr4pn2}, namely 4PM.} As discussed  in \cite{pmeft}, to isolate the potential modes we adapt to our EFT framework the powerful tools developed in \cite{zvi1,zvi2, Parra}. Notably, we make use of the methodology of differential equations using boundary conditions from the (static) limit $\gamma \to 1$ \cite{Parra}.\vskip 2pt  On the one hand, for diagrams $(c)$ and $(d)$ in Fig.~\ref{fig1}, only the $M^{(1,1)}_{n_1,n_2;\cdots}$ in \eqref{master} are needed, with $(n_1,n_2) \leq 0$, plus mirror images. These integrals, which contribute to the one-point function of a (boosted) Schwarzschild background, can be computed in the rest frame 
\beq u_1=(1,0,0,0)\,, \,\, u_2=(\gamma,\gamma\beta,0,0)\,, \label{rest}
\eeq
with $\beta\gamma = \sqrt{\gamma^2-1}$ \cite{pmeft}. At the end of the day, they turn into the same type that appear in the static limit of the PN expansion, see e.g.~\cite{nrgr2pn}. 
For diagrams ($e$), $(f)$ and $(g)$ in Fig.~\ref{fig1}, on the other hand, the  $M^{(1,2)}_{n_1 n_2;\cdots}$ are required instead, also with $(n_1,n_2) \leq 0$. Remarkably, the associated integrals for all these diagrams can be decomposed into a basis involving only the $M^{(1,2)}_{0 0;\cdots}$ subset~\cite{boot}. Furthermore, using integration by part (IBP) relationships \cite{Chetyrkin:1981qh, Tkachov:1981wb}, the contribution from diagrams $(e)$ and $(f)$ in Fig.~\ref{fig1} reduces to integrals with $i_3=0$. 
It is then straightforward to show that both diagrams vanish in $D=4$. (This is reminiscent of the fact that they do not enter at 2PN either \cite{nrgr2pn}.) Using the IBP relations and the aid of \texttt{FIRE6} \cite{Smirnov:2019qkx} and \texttt{LiteRed} \cite{Lee:2012cn}, as well as symmetry arguments, the calculation of the remaining (so-called $H$) diagram in Fig.\ref{fig1}~(g) is reduced to the following basis~\cite{boot} 
\beq
\left\{I_{11111}, I_{11211}, I_{01101},  I_{11011}, I_{00211}, I_{00112}, I_{00111} \right\}, \label{set}
\eeq
with $I_{i_1\cdots i_5} \equiv M^{(1,2)}_{00;i_1\cdots i_5}$. For the computation we follow~\cite{Henn:2013pwa} and various tools, e.g.~\texttt{epsilon} \cite{Prausa:2017ltv}, to construct a canonical basis $\vec h = \{h_{n=1\cdots 7}\}$ such that the velocity dependence is obtained via differential equations, 
\begin{align}\label{}
\partial_x\vec{h}(x,\epsilon) \,=\, \epsilon\, \mathbb{M}(x)\, \vec{h}(x,\epsilon)\,
\end{align}
with $\gamma = (x^2+1)/(2x)$, as advocated in \cite{Parra}. Because the set in \eqref{set} contains up to five (quadratic) propagators only, the associated boundary conditions in our case are then reduced to the same type of integrals that appear in the~PN regime at two loops ({\it Kite} diagrams, e.g.~\cite{nrgrG5}). It turns out only a handful contribute to the $H$ diagram in $D=4$, featuring the much anticipated factor of $\log x$ observed in~\cite{zvi1,zvi2,Parra}.\vskip 2pt 

To complete the derivation we have to include the iterations. Surprisingly, the set in \eqref{set} is (almost) sufficient for all the contributions. For instance, iterations involving the deflection due to Fig.\ref{fig1}~$(a)$ at LO order for the impulse due to Fig.~\ref{fig1}~$(b)$, and vice verse, follow~from \eqref{set}. Yet, for the deflection from Fig.\ref{fig1}~$(a)$ to NLO additional integrals are needed, resembling other (cut) topologies in \cite{zvi2,Parra}. In our case, we need the following~two:\footnote{In principle we find all $\pm i0$ combinations. Naively, due to the lack of `crossing' (e.g. $u_1 \to - u_1$) in the potential region, the connection between them is not obvious, see \cite{Parra}. Yet, we can show these integrals are related in~the static limit (see text). The upshot is that various $\pm i0$ choices differ by relative factors of $2$. (We thank Julio Parra-Martinez and Mao Zeng for discussions about this point.) These turn out to be crucial to ensure the cancellation of intermediate spurious infrared poles $\propto t^{-2\epsilon}/\epsilon^2$~\cite{boot}.} 
  \beq \{M^{(1,1)}_{11;11100}\, ,\,\, M^{(1,2)}_{11;11100}\} \,.\eeq 
Due to the presence of divergences, however, their computation is somewhat subtle. For the first one we can readily go to the rest frame in \eqref{rest} producing a $D-1$ integral. We then use the symmetrization described in~\cite{Parra}. Alternatively, it may be computed using the prescription in \cite{cheung,zvi1,zvi2} in the $u_2$-frame. Both can be adapted to all $\pm i0$ choices. The result is proportional to (twice) the standard one loop {\it bubble} integrals with static PN sources \cite{nrgr2pn}, although in $D-2$ dimensions. The same trick does not apply to the latter, but it can be easily incorporated into the canonical basis to obtain its $\gamma$-dependence. Yet, due to a divergence in the static limit, we need some care with the boundary condition. This is accounted for in the canonical basis by pulling out the relevant factor of $\beta$ (and $\epsilon$).  Once again we perform the integral in the rest frame, expand in small velocity and retain the leading term in $1/\beta$. In this limit, the $M^{(1,2)}_{11;\ldots}$ integral turns out to be equivalent (modulo different $\pm i0$ choices) to~the $M^{(1,1)}_{11;\ldots}$~counterpart.  We have checked all these relationships explicitly via a standard $\alpha$-parameterization~\cite{Smirnov}. At the end, as expected, the associated divergences cancel out in the final answer without subtractions.\vskip 2pt The~above steps culminate the derivation of the master integrals in the potential region via differential equations. Using various arguments, the boundary conditions are reduced to the master integrals that appear in the static limit of the PN expansion at the same loop order. See~\cite{boot} for a more detailed discussion.\vskip 2pt 

\emph{Scattering data.} The result for the impulse now follows from basic algebraic manipulations, and we arrive at
\begin{align}
\Delta^{\! (3)} p_1^\mu =\,&  \frac{G^3 b^\mu}{|b^2|^2}\Bigg(\frac{16m_1^2m_2^2(4\gamma^4-12\gamma^2-3)\Arcsinh\sqrt{\frac{\gamma-1}{2}}}{(\gamma^2-1)}\nn \\
&-\frac{4 m_1^2m_2^2\gamma(20\gamma^6-90\gamma^4+120\gamma^2-53)}{3(\gamma^2-1)^{5/2}}\nn \\
&-\frac{2m_1m_2(m_1^2+m_2^2)(16\gamma^6 {-} 32\gamma^4 {+} 16\gamma^2 {-} 1)}{(\gamma^2-1)^{5/2}}\Bigg)\nn \\
&+\frac{3\pi}{2}  \frac{\left(2\gamma^2-1\right)(5\gamma^2-1)}{(\gamma^2-1)^2} \frac{G^3M^2\mu}{|b^2|^{3/2}} \nn \\ 
& \quad\times\Big((\gamma m_2+m_1) u_2^\mu -(\gamma m_1+m_2) u_1^\mu \Big).\label{dp3}
\end{align}
The last term, which does not feature in the deflection angle at this order, is obtained from \eqref{onshell} and the result in~\cite{pmeft}. Hence, using \eqref{chi2}, the 1PM angle (cube) and the 2PM impulse along the velocities in \cite{pmeft}, we find 
\begin{align}\label{}
\frac{\chi^{(3)}_b}{\Gamma} =\,&  \frac{1}{(\gamma^2-1)^{3/2}}\Bigg[-\frac{4\nu}{3} \gamma\sqrt{\gamma^2-1}(14\gamma^2+25) \nn\\
&+ \frac{(64\gamma^6-120\gamma^4+60\gamma^2-5)(1+2\nu(\gamma-1))}{3(\gamma^2-1)^{3/2}}\nn\\
&-  8\nu (4\gamma^4-12\gamma^2-3)\Arcsinh\sqrt{\frac{\gamma-1}{2}}\,\Bigg]\,,\label{chi3}
\end{align}
which, using $\chi_j^{(3)} =(p_\infty/\mu)^3 \chi_b^{(3)} = \big(\sqrt{\gamma^2-1}/\Gamma\big)^3  \chi_b^{(3)}$, is in agreement with the derivation in \cite{zvi1,zvi2}, see also~\cite{Antonelli:2019ytb}. \vskip 2pt

\emph{B2B map.} The scattering data allows us to construct the (reduced) radial action \cite{paper1,paper2}
\beq
i_r =  \frac{p_\infty}{\sqrt{-p_\infty^2}}
\chi^{(1)}_j - j \left(1 + \frac{2}{\pi} \sum_{n=1}^{\infty}  \frac{\chi^{(2n)}_j}{(1-2n)j^{2n}}\right)  \,,\label{eq:ir}
\eeq
via analytic continuation to $\gamma <1$. As we discussed in \cite{paper1,paper2}, the natural power counting in $1/j$ in the PM expansion requires the (so far unknown) $\chi^{(4)}_j$ coefficient. The latter can be written, using the results in \cite{paper1,paper2}, as 
\beq
\chi^{(4)}_j = \frac{3\pi}{8M^4\mu^4}  \Big(P_1 P_3 + \frac{1}{2}P_2^2 + p_\infty^2 P_4\Big)\label{chi4}\,,
\eeq
with the $P_n$'s from the expansion of the CoM momentum
\beq
\bp^2 = p_\infty^2 + \sum_{n=1}^\infty P_n(E) \left(\frac{G}{r}\right)^n\label{pinfpm}\,.
\eeq
The $P_{n}$'s can also be obtained from the scattering angle, as described in~\cite{paper1,paper2}. For instance, inverting the relation 
\beq
\chi_j^{(3)} = \frac{1}{M^3\mu^3 p_\infty^3} \left(-\frac{P_1^3}{24} + p_\infty^2 \frac{P_1 P_2}{2} + p_\infty^4 P_3\right)\,, \label{chij3}
\eeq
together with \eqref{chi3} and the results in \cite{pmeft}, yields 
\beq
\begin{aligned}
&\frac{P_3}{M^3\mu^2} = \Bigg(\frac{18\gamma^2 - 1}{2\Gamma} +
 \frac{8\nu}{\Gamma}(3 {+} 12\gamma^2 {-} 4\gamma^4)\frac{\Arcsinh\sqrt{\frac{\gamma-1}{2}}}{\sqrt{\gamma^2-1}}+\\ 
&  \frac{\nu}{6\Gamma} \left(6 {-} 206 \gamma {-}108 \gamma^2 {-} 4\gamma^3 + \frac{18\Gamma(1-2\gamma^2)(1-5\gamma^2)}{(1+\Gamma)(1+\gamma)} \right)\Bigg)\,.\label{P3}
\end{aligned}
\eeq
This compact expression encodes all the information at 3PM order. It can be analytically continued to negative binding energies ($\gamma<1$) to derive observables for binary systems via the B2B map. Because of the factor~of $p_\infty^2$ in \eqref{chi4}, and since \eqref{pinfpm} has a well-defined static limit, the contribution in \eqref{eq:ir} from $P_4$ is subleading in the PN~expansion. This allows us to perform a consistent PN-truncation by keeping the $P_{n\leq 3}$ terms in \eqref{chi4} (ignoring also higher orders in $1/j$ which are PN-suppressed). This is carried out in detail in \cite{paper1,paper2}, and shown to agree with the literature in the overlapping regime of validity.\vskip 2pt   

\emph{Amplitude \& Hamiltonian.} It is instructive to use the B2B dictionary to also reconstruct both, the classical limit of the scattering amplitude as well as the Hamiltonian for the two-body system in the CoM (isotropic) frame. Using 
the relationship found in~\cite{paper1},
\beq
\bp^2 = p_\infty^2 + \frac{1}{2E} \int \dd^3\br  {\cal M}(p_\infty,\bq) e^{i\bq\cdot \br},
\eeq
we immediately read off from \eqref{P3} the (infrared-finite part of the) scattering amplitude in the classical limit, which agrees with the result in \cite{zvi2} (see Eq.\,(9.3)). For~the PM expansion of the Hamiltonian,
\beq
H(r,\bp^2) = \sum_i \frac{c_i(\bp^2)}{i!} \left(\frac{G}{r}\right)^i\,, \label{Ham}
\eeq
the coefficients can also be expressed iteratively in terms of the $P_n$'s in \eqref{pinfpm} \cite{paper1}. To~3PM order we find 
\begin{align}
    \frac{c_3(\bp)}{3!} = &-\frac{P_3(E)}{2E\xi}+\frac{(3\xi-1)P_2(E)P_1(E)}{4E^3\xi^3} \nonumber\\
    &+\frac{(P_2(E)P'_1(E)+P'_2(E)P_1(E))}{4E^2\xi^2} \nonumber\\
    &-\frac{(5\xi^2-5\xi+1)P_1^3(E)}{16E^5\xi^5}
    -\frac{(9\xi-3)P_1^2(E)P'_1(E)}{16E^4\xi^4} \nonumber\\
    &-\frac{P_1^2(E)P''_1(E)}{16E^3\xi^3}-\frac{P_1(E)(P'_1(E))^2}{8E^3\xi^3}\,,\label{c3}
\end{align}
where prime denotes a derivative with respect to $E$, and $\xi \equiv E_1E_2/(E_1+E_2)^2$. Inputting \eqref{P3}, and $P_{1,2}$ from~the 2PM results~\cite{pmeft}, we exactly reproduce the $c_{3}$ in \cite{zvi1,zvi2}. Notice, however, that the relevant PM information to compute observables through the B2B map is (more succinctly) encoded in \eqref{P3} at two loops, and ultimately the (yet to be computed) scattering angle at~4PM order. 

\vskip 8pt
{\it Conclusions}. Using the EFT approach and B2B dictionary  \cite{paper1,paper2,pmeft}, we derived the conservative dynamics for~non-spinning binary systems to 3PM order. Our~results, purely within the classical realm, are in perfect agreement with those reported in \cite{zvi1,zvi2}, thus removing the objections raised in \cite{damour3} against their validity. Even though, unlike the approach in \cite{zvi1,zvi2}, our derivation entails the use of Feynman diagram, because of the simplifications of the EFT/B2B framework just a handful are required (two of which are zero) at this order, see Fig.~\ref{fig1}. Moreover, only massless integrals appear and, as it was already illustrated in \cite{pmeft}, we do not encounter the (super-classical) infrared singularities which have, thus far, polluted the extraction of classical physics from the amplitudes program. By adapting to our EFT approach the methods in \cite{cheung,zvi1,zvi2,Parra}, we~found that the contribution from potential modes to the master integrals can be computed to all orders in velocities using differential equations (without the need of the PN-type resummations in \cite{zvi1,zvi2}). Remarkably, the boundary conditions are obtained from the knowledge of the same master integrals which appear in the static limit with PN sources to two loops, albeit in $D-1$ and $D-2$ dimensions. This implies that the PM dynamics can be bootstrapped from PN information (at least to NNLO). This is not surprising for the evaluation on the unperturbed trajectory, which serves as a {\it stationary} limit of the PM regime, but strikingly the same occurs for the iterations. Since master integrals for the PN expansion are known to four loops~\cite{nrgrG5}, bootstrapping integrals through differential equations could potentially give us up~to~the 5PM order.\vskip 2pt We note also that the infusion of data from outside of PN/PM schemes can further simplify the computation. For instance, the test-particle limit in a Schwarzschild background provides us the value of the $M^{(1,1)}$ master integrals in the iterations. In~turn, these are related to the $M^{(1,2)}$ family in the static limit. This would then allow us to read off their boundary condition directly from the test-body limit, and subsequently the entire velocity dependence with the differential equations. The fact that we get extra mileage from the probe limit is not surprising \cite{paper1}. 
What is remarkable, and more so due to the lack of crossing symmetry,\footnote{While the spurious infrared poles from the master integrals ultimately cancel out, crossing may be restored by implementing the zero-bin subtraction to remove the overlap with other `soft' regions, as with potential/radiation modes in the PN case~\cite{apparent,lamb}.} is the connection to ${\cal O}(\nu)$ corrections through the static limit and differential equations. 
Likewise, information from the gravitational self-force program \cite{gsf,adam}  may be also used to aid the calculation in the PM expansion, e.g.~\cite{damour1,damour2,Antonelli:2019ytb,damour3n,damour3,bini1,bini2,Vines:2018gqi,justin2,binit}. Irrespectively of the weapon of choice, the B2B dictionary \cite{paper1,paper2} is imploring us to continue to {\em even} higher orders. The~derivation of the needed 4PM scattering angle is ongoing in the EFT approach, which we have demonstrated here is a powerful framework, not only for PN calculations \cite{nrgr,walterLH,foffa,iragrg,grg13,review}, but also in the PM regime \cite{pmeft,tidalpm}. 

\vskip 4pt
{\it Acknowledgements}. We thank  Babis Anastasiou, Zvi Bern, Clifford Cheung, Lance Dixon, Claude Duhr,~Julio Parra-Martinez, Radu Roiban, Chia-Hsien Shen, Mikhail Solon, Gang Yang and Mao Zeng for useful discussions. We are grateful to Julio Parra-Martinez and~Mao Zeng for helpful comments on the integration in the potential region.  R.A.P. acknowledges financial support from the ERC Consolidator Grant ``Precision Gravity: From the LHC to LISA"  provided by the European Research Council (ERC) under the European Union's H2020 research and innovation programme (grant agreement No.\,817791). Z.L.~and R.A.P.~are also supported by the Deutsche Forschungsgemeinschaft (DFG) under Germany's Excellence Strategy (EXC 2121) `Quantum Universe' (390833306). G.K. is supported by the Knut and Alice Wallenberg Foundation under grant KAW 2018.0441, and in part by the US DoE under contract DE-AC02-76SF00515.

\bibliography{ref3PM}

\end{document}